\documentstyle[aps,prl,multicol]{revtex}
\DeclareSymbolFont{AMSb}{U}{msb}{m}{n}
\DeclareSymbolFontAlphabet{\mathbb}{AMSb}
\newcommand{\complex}{\kern.1em{\raise.47ex\hbox{
            $\scriptscriptstyle |$}}\kern-.40em{\rm C}}

\newcommand{\C}{\complex}
\newcommand{\CP}{{{\mathbb{CP}}^1}}
\newcommand{\Tr}{{\hbox{Tr}}}
\newcommand{\be}{\begin{equation}}
\newcommand{\ee}{\end{equation}}
\newcommand{\bearray}{\begin{eqnarray}}
\newcommand{\eearray}{\end{eqnarray}}
\title{Geometry, Statistics and Asymptotics of  Quantum Pumps}
\begin{document}
\input epsf
\tightenlines
\author{ J.~E.~Avron ${}^{(a)}$, A. Elgart ${}^{(a)}$, G.M. Graf ${}^{(b)}$
and L. Sadun ${}^{(a,c)}$}
\address{${}^{(a)}$ Department of Physics, Technion, 32000 Haifa,
Israel}
\address{${}^{(b)}$ Theoretische Physik, ETH -H\"onggerberg, 8093 Z\"urich,
Switzerland}
\address{${}^{(c)}$ Department of Mathematics, University of Texas, Austin Texas 78712, USA}
 \draft
\maketitle
\begin{abstract}
We give a pedestrian derivation of a formula of B\"uttiker et.\
al.\ (BPT) relating the adiabatically pumped current to the $S$
matrix and its (time) derivatives. We relate the charge in BPT to
Berry's phase and the corresponding Brouwer pumping formula to
curvature. As applications we derive explicit formulas for the
joint probability density of pumping and conductance when the $S$
matrix is uniformly distributed; and derive a new formula that
describes hard pumping when the $S$ matrix is periodic in the
driving parameters.

\end{abstract}
\begin{multicols}{2}
\narrowtext
 Brouwer
\cite{brouwer}, and Aleiner et.\ al.\ \cite{aleiner}, building on
results of B\"uttiker, Pretre and Thomas (BPT) \cite{bpt}, pointed
out that  adiabatic scattering theory leads to a geometric
description of charge transport in mesoscopic quantum pumps. Some
of these works, and certainly our own work, was motivated by
experimental results of Switkes et. al. \cite{marcus} on such
pumps.

In this article we examine the formula of BPT \cite{bpt}, which
relates adiabatic charge transport to the $S$ matrix and its
(time) derivatives, in the special case of single-channel
scattering. We show that the formula admits a simple
interpretation in terms of three basic processes at the Fermi
energy. Two of these are dissipative and non-quantized. The third
integrates to zero for any cyclic variation in the system.

Next, we describe the geometric significance of BPT and relate it
to Berry's phase \cite{berry}. It follows that the pumping formula
of Brouwer \cite{brouwer} can be interpreted as curvature and is
formally identical to the adiabatic curvature\cite{berry}. In
spite of the interesting geometry the topological aspects of
pumping are trivial. In particular, we prove that all Chern
numbers associated to the Brouwer formula are identically zero.

We proceed with two applications. First we give an elementary and
explicit derivation of the joint probability density for pumping
and conductance. This problem was studied in \cite{brouwer}.
Brouwer's results go beyond ours as he also calculates the tails
of the distributions and we don't. On the other hand, parts of his
results are numerical, and they are certainly not elementary.
Finally, we calculate, for the first time, the asymptotics of hard
pumping for $S$ matrices that depend periodically on two
parameters. If the system traverses a circle of radius $R$ in
parameter space, with $R$ large, then the amount of charge
transported is  order $\sqrt{R}$, multiplied by a quasi-periodic
(oscillatory) function of $R$ leading to ergodic behavior.

We shall use units where $e=m=\hbar=1$, so the electron charge is
$-1$ and the quantum of conductance is $e^2/h = {1 \over 2\pi}$.
The mutual Coulombic interaction of the electrons is disregarded.

{\bf The BPT formula:}
 Consider a scatterer connected to
 leads that terminate at electron reservoirs. All the reservoirs are
initially at the same chemical potential and at zero temperature.
The scatterer is described by its (on-shell) $S$ matrix, which, in
the case of $n$ channels is an $n\times n$  matrix  parameterized
by the energy and other parameters associated with the adiabatic
driving of the system (e.g. gate voltages and magnetic fields).

The BPT formula \cite{bpt} says that the charge $dq_\ell$ entering
the scatterer from the $\ell$-th lead
due to an adiabatic variation of $S$  is
\be
d{ q}_\ell = {i  \over 2 \pi}\, \Tr \left (Q_\ell \,dS
S^\dagger \right ),
\label{dQ} \ee
where $Q_\ell$ is a projection on the channels in the $\ell$-th lead,
and the $S$ matrix is evaluated at the Fermi energy. In the special case of two
leads, each lead carrying a single channel,
\be
S = \pmatrix{r & t' \cr t & r'},\quad Q_\ell=\pmatrix{1&0\cr 0&0}
\label{whatsS} \ee where $r,\ (r') $ and $t,\ (t')$ are the
reflection and transmission coefficients from the left (right) and
$Q_\ell$ projects on the left lead. In this case Eq.~(\ref{dQ}),
for the charge entering through the left lead, reduces to
\be\label{bpt} 2\pi\,d{ q}_\ell={i}\,(\bar r dr + \bar t'd
t').\label{dq} \ee We shall present an elementary derivation of
(\ref{dq}).

{\bf Derivation:} Every unitary $2 \times 2$ matrix can be
expressed in the form: \be S = e^{i\gamma} \pmatrix{\cos(\theta)
e^{i \alpha} & i \sin(\theta) e^{-i\phi} \cr i \sin(\theta)
e^{i\phi} & \cos(\theta) e^{-i \alpha}}, \label{S1} \ee where
$0\le \alpha,\phi<2\pi,\ 0\le\gamma<\pi$ and $0\le\theta\le\pi/2$.
In terms of these parameters, Eq.~(\ref{bpt}) reads
 \be 2 \pi d{
q}_\ell=-\cos^2(\theta) d\alpha+\sin^2(\theta)d\phi -
d\gamma.\label{dq2} \ee

The basic strategy of
our derivation of Eq.~(\ref{dq2}) is to find processes that vary
each of the parameters in turn, and keep track of how much current
is generated by each process.  An underlying assumption is that
current depends only on $S(k_F)$ and $\dot S(k_F)$, so that
processes that give rise to the same change in the $S$
matrix also give rise to the same current. Because we do not prove
this assertion, our derivation cannot be considered a complete proof.

\begin{figure}
\centerline{\epsfxsize=2in \epsfbox{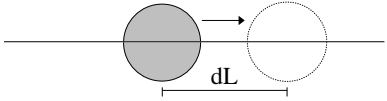}} \caption{Moving
the scatterer changes $\alpha \to  \alpha + 2k_F d L$}
\end{figure}

\begin{figure}
\centerline{\epsfxsize=2in \epsfbox{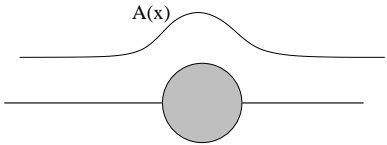}}
\caption{Applying a vector potential changes $\phi \to  \phi -
\int A$}
\end{figure}

We understand the four parameters as follows. (See figures.) The parameter
$\alpha$ is associated with translations: Translating the
scatterer a distance $dL=d \alpha/2k_F$ to the right multiplies
$r,\ (r')$ by $e^{id \alpha},\ (e^{-id \alpha}) $, and leaves $t$ and
$t'$ unchanged.  The parameter $\phi$ is associated with a vector
potential $A$ near the scatterer. This induces a phase
shift $d\phi = -\int A$ across the scatterer, and multiplies
$t,\ (t')$ by $e^{i d \phi},\
(e^{-i d \phi}) $, while leaving $r$ and $r'$ unchanged. The parameter
$\theta$ determines the conductance of the system: $g= |t|^2/2\pi
= \sin^2(\theta)/2\pi$. Finally, $\gamma = (-i/2)\log\det S$ is
related, by Krein's spectral shift \cite{krein}, to the number of
electrons trapped in the scatterer. As a consequence, for any closed
path in the space of Hamiltonians  $\oint d\gamma=0$.

To determine the effect of changing $\alpha$ we imagine a process
that changes $\alpha$ and leaves the other parameters fixed,
namely translating the scatterer a distance $dL = d\alpha/2k_F$
to the right. The scatterer passes through a fraction $|t|^2$ of
the $k_F d L/\pi = d\alpha/2\pi$ electrons that occupy the
region of size $d L$, and pushes the remaining $|r|^2
d\alpha/2\pi $ electrons forward. Thus
\be 2 \pi dq = -
\cos^2(\theta) d\alpha. \label{dalpha} \ee This result can be
obtained less heuristically, \cite{ap}, by working in the
reference frame of the moving scatterer and integrating the
contribution of each wave number from 0 to $k_F$. From this one
also sees \cite{ap} that the rate of dissipation at the
reservoirs, $P$, is quadratic in the current $I$, with a coefficient
that depends on the dispersion relation.
If the dispersion relation is quadratic, then
\be
P= 2 \pi I^2 / |r(k_F)|^2.
\ee

 To change
$\phi$, we vary the vector potential. This induces an EMF of
strength $ -\int \dot A  = \dot\phi$. The current is simply the
voltage times the Landauer conductance $|t|^2/2\pi$
\cite{landauer}. Integrating over time gives
\be
2 \pi dq = \sin^2(\theta) d \phi. \label{dphi} \ee A current $I$
 then dissipates energy at the reservoirs at a rate
\be P= 2 \pi I^2 /
|t(k_F)|^2.
\ee

To understand the effect of changing $\theta$ and $\gamma$, we
first suppose our scatterer is right-left symmetric, so $r=r'$ and
$t=t'$.   Then changes in $\theta$ and $\gamma$ would draw equal
amounts of charge to the scatterer from the left and right leads.
The charge that accumulates on the scatterer is given by Krein's
spectral shift \cite{krein}. The charge coming from the left is
half this, namely \cite{ap}:
\be
2 \pi dq={2 \pi i \over 4\pi} \, d\log\det S = - d \gamma.
\label{dgamma} \ee Since every $S$ matrix can be obtained by
translating and adding a vector potential to a right-left
symmetric scatterer, the formula (\ref{dgamma}) applies to all
possible $S$-matrices. The formula (\ref{dgamma}) applies to all
possible scatterers, symmetric or not.

Combining (\ref{dalpha}),
(\ref{dphi}) and (\ref{dgamma}), gives BPT, Eq.~(\ref{dq2}).

The effect of changing $\gamma$ integrates to zero on a closed
loop. Changing $\theta$ does not give any transport at all. Thus,
only changes in $\alpha$ and $\phi$ contribute to the net
transport of a quantum pump.  These are dissipative processes,
with the rate of energy dissipation $P$ (in the reservoirs) that
is bounded from below by $2 \pi I^2$.
This is contrary to the assertions of \cite{aleiner}, who claimed that the charge
transport is the sum of a quantized non-dissipative term and a
dissipative term that is not quantized.

{\bf Geometrical interpretation:} ${\cal A}=-2\pi \,dq$  is the
1-form (vector potential) associated with Berry's phase\cite{berry}. If we define the
unit spinor $| \psi \rangle = \pmatrix{r \cr t'}$ then
\be
{\cal A} =
{-i} \langle \psi | d \psi \rangle. \label{BerrydQ}
\ee
%
%
%
The set of all spinors $| \psi \rangle$ is a 3-sphere (since $|r|^2+|t'|^2=1$), while
the set of all ratios $r/t'$ is the projective space $\CP = \C + \{\infty\}
\simeq S^2$.
The natural map between them, namely $| \psi \rangle \to r/t'$,
is called the Hopf fibration, and ${\cal A}$ is called the
 ``global angular form''
of this fibration.

 To compute the charge transported by a closed cycle $C$
in parameter space, we can either integrate the 1-form $\cal A$
around $C$, or (by Stokes' theorem) integrate the exterior
derivative (curl) $\Omega= d \cal A$ over a disk $D$ whose
boundary is $C$. $\Omega$ is the curvature  2-form of Brouwer
\cite{brouwer}: \bearray \Omega & = &{-i} \langle d\psi | d
\psi\rangle =  -i{ d\bar z\wedge  dz \over (1+|z|^2)^2},
\label{Omega} \eearray where $z=r/t'$. The expression $-i \langle
d\psi | d \psi\rangle$ is formally identical to the adiabatic
(Berry's) curvature that appears also in the quantum Hall effect
\cite{ThoulessBOOK}.

In the last expression one sees that the curvature sees only the
ratio $z=r/t'$, and not $r$ and $t'$ separately.
The curvature $\Omega$ is the $U(2)$-invariant area form on $\CP$, and its integral
over all of $\CP$ is $2\pi$. $\Omega$ is also the curvature of the
Hopf fibration.

In the study of non-dissipative quantum transport, Chern numbers
play a role.  These are
topological invariants that equal the integral of the curvature
over closed surfaces in parameter space. In the
context of adiabatic scattering, however, all Chern numbers are
zero. The vector bundle over parameter space is
topologically trivial, and the vector $(r, t')$ gives a section of
this bundle.

These geometrical constructions generalize to systems with $n$
incoming and $m$ outgoing channels, \cite{ap}.  The first $n$ rows
of $S$ span an $n$-dimensional complex subspace of $\C^{n+m}$. The
space of all such subspaces, called a Grassmannian, has a
naturally defined 2-form, called the K\"ahler form
\cite{kobayashi}. Up to a constant factor, the Brouwer 2-form
equals the K\"ahler form. In addition, there is a canonically
defined line bundle over this Grassmannian, and ${\cal A}$ equals
the global angular form for this bundle.


{\bf Statistics of weak pumping:} Next we consider how a random
scatterer transports charge when two parameters are varied gently
and cyclically.  More precisely, we consider the charge
transported by moving along the circle $X_1 = \epsilon
\cos(\tau)$, $X_2 =\epsilon \sin(\tau)$ in parameter space. If
$\epsilon$ is small, then the charge transport is close to $-
\frac{\pi \epsilon^2}{2\pi}\,\Omega(\partial_1, \partial_2)$,
evaluated at the origin, where $\partial_j$ are the tangent
vectors associated with the parameters $X_j$. The vectors
$\partial_j$ map to random vectors on $U(2)$, which we assume to
be Gaussian with covariance $C$.  The problem is then to
understand the possible values of the curvature $\Omega$ applied
to two random vectors.

To do this, we first need to understand the
statistics of 2-forms applied to pairs of random vectors, and
to understand the geometry of the group $U(2)$.

Take two random vectors in ${\mathbb R}^2$, and see how much area
they span. By random vectors we mean independent, identically
distributed Gaussian random vectors whose components $X_j$ have
the covariance $\langle X_i X_j \rangle = C \delta_{ij}$. The area
$A$ is distributed as a 2-sided exponential: \be dP(A) = {1\over
2C}\, e^{-|A|/C}\, dA.\label{exp} \ee This is seen as follows. If
the two vectors are $X$ and $Y$, then the area is $X_1 Y_2 - X_2
Y_1$. $X_1 Y_2$ and $-X_2 Y_1$ are independent random variables,
and a calculation shows that their characteristic function is
$1/\sqrt{k^2 C^2+1}$. Their sum is a random variable with
characteristic function $(k^2C^2 + 1)^{-1}$, and so exponentially
distributed.

We parameterize the group $U(2)$ by the angles $(\alpha, \gamma,
\phi, \theta)$, as in (\ref{S1}). A standard, bi-invariant metric
on $U(2)$ is \bearray\frac{1}{2}\,Tr(dS^*\otimes
dS)&=&(d\gamma)^2+ \cos^2\theta\, (d\alpha)^2+\cr &+&\sin^2\theta\,
(d\phi)^2+(d\theta)^2. \eearray In this metric the vectors
$\partial_i$ are orthogonal but not orthonormal. Unit tangent
vectors are \be e_\gamma=\partial_\gamma, \ e_\alpha=
\frac{1}{\cos\theta}\,
\partial_\alpha,\ e_\phi=\frac{1}{\sin\theta}\,
\partial_\phi,\ e_\theta=\partial_\theta. \ee The volume form is
$\sin(\theta)\cos(\theta)\, d\alpha\wedge d\gamma\wedge
d\phi\wedge d\theta.$ The curvature 2-form, from
Eq.~(\ref{Omega}), is \be\label{omega} \Omega = {-2}
\sin(\theta)\cos(\theta)\, d\theta \wedge(d\alpha + d\phi)
\label{Omega2}. \ee

A scattering matrix is time reversal invariant if and only if $t =
t'$. The space of time-reversal matrices is parameterized exactly
as before, only now with $\phi$ identically zero.  The volume form
for the metric inherited from $U(2)$ is $\cos(\theta)
d\alpha\wedge d\gamma\wedge d\theta, \label{volume2}$ and the
curvature form is now $\Omega = {-2} \sin(\theta)\cos(\theta)
d\theta \wedge d\alpha$.

We are now prepared to compute the statistics of weak pumping,
assuming that the $S$ matrix is uniformly distributed and that the
tangent vectors to the space of $S$ matrices are Gaussian random
variables. This problem was studied by Brouwer \cite{brouwer}, in
the framework of random matrix theory, by which we mean that
Brouwer posits an a-priori measure on the space of {\em
Hamiltonians}. Random matrix theory is more powerful, in that the
distribution of tangent vectors is fixed by the theory. The price
one pays is that the analysis is also far from elementary and the
results are, in part, only numerical.

For systems without time reversal symmetry, random matrix theory
posits that the $S$ matrix is distributed on $U(2)$ with a uniform
measure. Since the
conductance $g$  is $g\propto |t|^2 = \sin^2\theta$ we have
that $dg\propto \sin\theta\cos\theta d\theta$, proportional to the
volume form: The conductance $g$ is therefore uniformly
distributed.

A random tangent vector to $U(2)$ is $X= X_\theta e_\theta+
X_\alpha e_\alpha+X_\phi e_\phi+X_\gamma e_\gamma,$ where $X_j$
are Gaussians with $\langle X_jX_k\rangle=C \delta_{jk}$. The
curvature associated with two random tangent vectors $X,\ Y$ is,
by Eq.~(\ref{omega}) \be \Omega(X,Y)= -2\Big( X_\theta \, W_\theta
-Y_\theta\,Z_\theta\Big), \ee where $W_\theta=\sin\theta\,
Y_\alpha+\cos\theta \,Y_\phi$ and
$Z_\theta=\sin\theta\,X_\alpha+\cos\theta\, X_\phi.$
The variables $W_\theta$ and $Z_\theta$ are independent,
each with variance $C$. From
Eq.~(\ref{exp}), the distribution of the curvature is exponential
and independent of $|t|$.  The joint distribution of curvature,
$\omega$, and conductance, $g= {1\over 2\pi}\, |t|^2$ is given by
the probability density
\be {\pi \over 2 C}\ e^{-|\omega|/2C}
d\omega\, dg  \label{NoTRI2}
\ee
with $\omega$ ranging from
$-\infty$ to $\infty$ and $g$ from 0 to $1\over 2\pi$.

For systems with time reversal symmetry, the $S$ matrix is
uniformly distributed on the $t=t'$ submanifold, with the metric
inherited from $U(2)$. The tangent vectors are now Gaussian random
variables of the form $ X= X_\theta e_\theta+ X_\alpha
e_\alpha+X_\gamma e_\gamma,$ and the curvature is now
\be
\Omega(X,Y)= -2\sin\theta\,\Big( X_\theta \,
Y_\alpha-Y_\theta\,X_\alpha\Big).\ee Since the curvature depends
on $\theta$ the curvature and the conductance are  correlated. The
volume form indicates that $\sqrt g$, and not $g$, is uniformly
distributed. This favors insulators. The joint distribution for
curvature and conductance is \be {1 \over 4 \sqrt{g} C}\,
e^{-|\omega|/2C\sqrt{2\pi g}}\, d\omega\, d(\sqrt g).
 \label{TRI2}
\ee This formula says that, statistically, good pumps tend to be
good conductors;  $\omega/\sqrt{g}$, rather than $\omega$ itself,
is independent of $g$.

We have assumed, so far,  that the variance $C$ is a constant.
There is no reason for this and it is natural to let $C$ itself be
a random variable. Given a probability distribution for the
covariance, $d\mu(C)$, one integrates the formulas (\ref{NoTRI2})
and (\ref{TRI2}) over $C$. One sees, by inspection, that in the
absence (presence) of time reversal symmetry, $\omega$
($\omega/\sqrt{g}$) is independent of $g$. Furthermore, the
distribution of $\omega$ after integrating over $g$ is smooth away
from $\omega=0$, but has a discontinuity in derivative (log
divergence) at $\omega=0$. In these qualitative features, our
results agree with Brouwer's. However, the tails of the
distribution for large pumping may depend on the tail of
$d\mu(C)$; While (\ref{NoTRI2}) and (\ref{TRI2}) have
exponentially small tails, power law tails  $d\mu(C)$ will lead to
power law in the tails in $\omega$.  Since we do not determine
$d\mu(C)$ we can not determine the tails. Using random matrix
theory  Brouwer determined the power decay in $\omega$
\cite{brouwer}.

{\bf Hard Pumping:} Finally, we consider what happens for hard
pumping. Here one can no longer evaluate the curvature at a point
and multiply by the area.  One needs to honestly integrate the
curvature. Hard pumping was addressed by \cite{aleiner} who
studied it in the context of random matrix theory and showed,
using rather involved diagrammatic techniques, that  pumping scales
like the root of the perimeter. Here we shall describe a
complementary, elementary result that holds provided the $S$ matrix
is a periodic function of the parameters. This is the case, for
example, when the pumping is driven by two Aharonov-Bohm fluxes.

With $S(x,y)$ periodic in the driving parameters $x$ and $y$, so
is the curvature $\Omega(x,y)=\sum
\hat\Omega_{mn}\,e^{i(mx+ny)}.$  Since the global angular form is
also periodic, $\hat\Omega_{00}=0$.

The integral $\int_{|x|<R} \Omega$ of the curvature on a large
disc of radius $R$ is, to leading order, \be\label{gauss}
\sqrt{8\pi R}\,\sum \,\frac{\hat\Omega_{nm}}{N(n,m)^{3/2}}\,
\sin\left(N(n,m)\,R-\frac{\pi}{4}\right), \ee where
$N(n,m)=\sqrt{n^2+m^2}$. The charge transported in a cycle is
proportional to the square root of the perimeter (or the fourth
root of the area) times a quasiperiodic function of $R$. This
follows from the evaluation of the elementary integral
$\int_{|x|<R} dx\,dy\, e^{i(nx+my)}$, which equals a Bessel
function whose large-$R$ asymptotic behavior is $\sqrt{\frac{8\pi
R}{N^3}}\,\sin(NR-  \frac{\pi}{4}).$ From Eq.~(\ref{gauss}) one can
determine the probability distribution for charge transport,
(viewed as a random variable with uniform distribution on the
radius $R$). Eq.~(\ref{gauss}) turns out to be closely related to a
celebrated problem in ergodic number theory: The Gauss circle
problem \cite{bleher}.

This result does not directly apply to the pump studied by Switkes
\cite{marcus}, because the parameters they vary do not have built
in periodicity. Nevertheless, it illustrates two features of pumps
that have been observed experimentally. The first is that hard
driving transports a lot of charge, with scaling that is
sublinear, as in a random process, and the second that the
directionality of hard pumping is essentially unpredictable.

We thank A. Kamenev for extremely valuable insights and A.
Auerbach, P. Brouwer and C. Marcus for useful correspondences, and
U. Sivan for a careful reading of the manuscript, and valuable
suggestions. This research was supported in part by the Israel
Science Foundation, the Fund for Promotion of Research at the
Technion, the DFG, the National Science Foundation and the Texas
Advanced Research Program.

\end{multicols}

\end {document}